\documentclass[12pt]{article}
\renewcommand{\d}{{\rm d}}
\newcommand{\bl}{\mbox{\boldmath$l$}}

\newcommand{\pmtau}{\,^{\pm}\!\tau}
\newcommand{\pmpi}{\,^{\pm}\!\pi}
\newcommand{\pmbpi}{\,^{\pm}\!\mbox{\boldmath$\pi$}}

\newcommand{\pmbSig}{\,^{\pm}\!\mbox{\boldmath$\Sigma$}}
\newcommand{\pmSig}{\,^{\pm}\!\Sigma}
\newcommand{\pmbv}{\,^{\pm}\!\mbox{\boldmath$v$}}
\newcommand{\pmbw}{\,^{\pm}\!\mbox{\boldmath$w$}}
\newcommand{\pbv}{\,^{+}\!\mbox{\boldmath$v$}}
\newcommand{\mbv}{\,^{-}\!\mbox{\boldmath$v$}}
\newcommand{\pmv}{\,^{\pm}\!v}
\newcommand{\pmw}{\,^{\pm}\!w}
\newcommand{\bphi}{\mbox{\boldmath$\phi$}}

\newcommand{\sbl}{\mbox{\scriptsize\boldmath$l$}}
\newcommand{\br}{\mbox{\boldmath$r$}}
\newcommand{\sdef}{\stackrel{\rm def}{=}}
\newcommand{\ve}{\varepsilon}
\newcommand{\R}{\bar{R}}
\renewcommand{\t}{\textstyle}
\newcommand{\pmbtau}{\,^{\pm}\!\mbox{\boldmath$\tau$}}

\newcommand{\pbpi}{\,^+\!\mbox{\boldmath$\pi$}}
\newcommand{\mbpi}{\,^-\!\mbox{\boldmath$\pi$}}

\newcommand{\bn}{\mbox{\boldmath$n$}}
\newcommand{\pbtau}{\,^+\!\mbox{\boldmath$\tau$}}
\newcommand{\mbtau}{\,^-\!\mbox{\boldmath$\tau$}}
\newcommand{\at}{|\mbox{\boldmath$\tau$}|}
\newcommand{\sbtau}{\mbox{\scriptsize\boldmath$\tau$}}
\newcommand{\sbphi}{\mbox{\scriptsize\boldmath$\phi$}}
\newcommand{\btau}{\mbox{\boldmath$\tau$}}
\newcommand{\pmbphi}{\,^{\pm}\!\mbox{\boldmath$\phi$}}
\newcommand{\sat}{|\mbox{\scriptsize\boldmath$\tau$}|}

\newcommand{\ch}{{\rm ch}}
\newcommand{\sh}{{\rm sh}}

\begin{document}

\title{Modification of quantum measure in area tensor Regge calculus and positivity}
\author{V.M. Khatsymovsky \\
 {\em Budker Institute of Nuclear Physics} \\ {\em
 Novosibirsk,
 630090,
 Russia}
\\ {\em E-mail address: khatsym@inp.nsk.su}}
\date{}
\maketitle
\begin{abstract}
A comparative analysis of the versions of quantum measure in the area tensor Regge
calculus is performed on the simplest configurations of the system. The quantum
measure is constructed in such the way that it reduces to the Feynman path integral
describing canonical quantisation if the continuous limit along any of the
coordina\-tes is taken. As we have found earlier, it is possible to implement also the
correspon\-dence principle (proportionality of the Lorentzian (Euclidean) measure to
$e^{iS}$ ($e^{-S}$), $S$ being the action). For that a certain kind of the connection
representation of the Regge action should be used, namely, as a sum of independent
contributions of selfdual and antiselfdual sectors (that is, effectively 3-dimensional
ones). There are two such representations, the (anti)selfdual connecti\-ons being
SU(2) or SO(3) rotation matrices according to the two ways of decomposing full SO(4)
group, as SU(2) $\!\times\!$ SU(2) or SO(3) $\!\times\!$ SO(3). The measure from SU(2)
rotations although positive on physical surface violates positivity outside this
surface in the general configuration space of arbitrary independent area tensors. The
measure based on SO(3) rotations is expected to be positive in this general
configuration space on condition that the scale of area tensors considered as
parameters is bounded from above by the value of the order of Plank unit.
\end{abstract}
\newpage
In our previous work \cite{Kha1} we have constructed the quantum measure in area
tensor Regge calculus with the following property. Whatever coordinate $t$ is chosen
along which the continu\-ous limit is taken, the resulting (properly defined)
continuous limit of the quantum measure is the Feynman path integral corresponding to
the canonical quantisation of the resulting system with continuous time $t$. The
possibility for such measure to exist is specific for the area tensor Regge calculus.
The latter reminds in this respect the 3-dimensional Regge calculus for which the same
problem has been solved in our earlier work \cite{Kha2}.

In three dimensions the constructed completely discrete measure takes the form
\begin{equation}                                                                    
\label{d-mu-3}%
\d\mu_3 = \exp\left (i\sum_{\sigma^1}\bl_{\sigma^1}*R_{\sigma^1}(\Omega)\right
)\prod_{\sigma^1\not\in{\cal F}}\d^3\bl_{\sigma^1}\prod_{\sigma^2}{\cal
D}\Omega_{\sigma^2}, ~~~ \bl*R\stackrel{\rm def}{=}{1\over 2}l^aR^{bc}\epsilon_{abc}.
\end{equation}

\noindent Here the field variables are the vectors $\bl_{\sigma^1}$ on the edges
$\sigma^1$ (the 1-dimensional simplices $\sigma^1$) and SO(3) connection matrices
$\Omega_{\sigma^2}$ on the triangles $\sigma^2$ (the 2-simplices $\sigma^2$). The
curvature matrix $R_{\sigma^1}(\Omega)$ is the (ordered along the path enclosing the
edge $\sigma^1$) product of the matrices $\Omega^{\pm 1}_{\sigma^2}$ for the triangles
$\sigma^2$ containing $\sigma^1$. The ${\cal F}$ is some set of edges arranged in
non-intersecting and non-self-intersecting broken lines passing through each vertex of
the manifold. Considered is the Euclidean signature case, and the integral in
(\ref{d-mu-3}) is defined by means of rotation of the integration contours in the
complex plane, $\bl$ $\to$ $-i\bl$ (to simplify notations, we do not show in
(\ref{d-mu-3}), but imply that this rotation is made only for those $\bl_{\sigma^1}$
which are the integration dummy variables, not for $\sigma^1$ $\in$ ${\cal F}$). The
${\cal D}\Omega$ is the Haar measure on SO(3). Arising it in the canonical approach is
connected with the structure of the Lagrangian originating from the Regge action in
the continuous time limit and resulting in the set of constraints which turn out to
coincide with those proposed for the 3-dimensional gravity by Waelbroeck \cite{Wael}
from symmetry considerations.

In four dimensions the result for the Euclidean measure being applied to arbitrary
function on the set of area tensors $\pi$ and connection matrices $\Omega$ reads
\begin{eqnarray}                                                                    
\label{VEV2}%
<\Psi (\{\pi\},\{\Omega\})> & = & \int{\Psi (-i\{\pi\}, \{\Omega\})\exp{\left (-\!
\sum_{\stackrel{t-{\rm like}}{\sigma^2}}{\tau _{\sigma^2}\circ
R_{\sigma^2}(\Omega)}\right )}}\nonumber\\
 & & \hspace{-20mm} \exp{\left (i
\!\sum_{\stackrel{\stackrel{\rm not}{t-{\rm like}}}{\sigma^2}} {\pi_{\sigma^2}\circ
R_{\sigma^2}(\Omega)}\right )}\prod_{\stackrel{\stackrel{\rm
 not}{t-{\rm like}}}{\sigma^2}}{\rm d}^6
\pi_{\sigma^2}\prod_{\sigma^3}{{\cal D}\Omega_{\sigma^3}} \nonumber\\ & \equiv &
\int{\Psi (-i\{\pi\},\{\Omega\}){\rm d} \mu_{\rm area}(-i\{\pi\},\{\Omega\})}, ~~~
\pi\circ R \stackrel{\rm def}{=}{1\over 2}\pi_{ab}R^{ab}.
\end{eqnarray}

\noindent We define tensors on the triangles $v^{ab}_{\sigma^2}$ (analogs of the
bivectors $\epsilon^{ab}{}_{cd}l^c_1l^d_2/2$ in ordinary Regge calculus formed by the
pairs of vectors $l^a_1$, $l^a_2$), now independent variables, and SO(4) connections
on the tetrahedra $\Omega_{\sigma^3}$. Here analog of the set ${\cal F}$ in
(\ref{d-mu-3}) consisting of the triangles $\sigma^2$ integration over tensors of
which is absent is specified. For that the regular method of constructing the
4-dimensional Regge manifold from the 3-dimensional Regge manifolds used in
\cite{Kha1} (analogous to that proposed in \cite{MisThoWhe}) is considered, these
3-dimensional manifolds usually being referred to as {\it the leaves of the
foliation}. A possible choice for the triangles of ${\cal F}$ are the $t$-like ones
where $t$ labels the sections, that is, those triangles one of edges of which is
located along the line of the coordinate $t$. Besides that, there are also the leaf
triangles completely contained in the 3-dimensional leaves and diagonal ones which are
neither leaf nor $t$-like triangles. Correspondingly, the tensors $v_{\sigma^2}$ are
divided into those of the $t$-like triangles $\tau_{\sigma^2}$ and of the leaf and
diagonal ones, $\pi_{\sigma^2}$. The integration is to be performed over only the
latter tensors, and the rotation of the integration contours is made just for them,
$\pi_{\sigma^2}$ $\to$ $-i\pi_{\sigma^2}$.

Another form of (\ref{VEV2}) convenient for calculation follows upon formal splitting
matrices into selfdual and antiselfdual parts,
\begin{eqnarray}                                                                    
\label{d-mu+-}%
{\rm d} \mu_{\rm area} & = & {\rm d} \,^+\!\mu_{\rm area}{\rm d} \,^-\!\mu_{\rm area},
\nonumber\\\lefteqn{{\rm d} \,^{\pm}\!\mu_{\rm area}(-i\{\pi\},\{\Omega\}) =
\exp{\left (-\! \sum_{\stackrel{t-{\rm like}}{\sigma^2}}{\pmtau _{\sigma^2}\circ
R_{\sigma^2}(\,^{\pm}\!\Omega)}\right )}}\nonumber\\
 & & \hspace{-15mm} \exp{\left (i
\!\sum_{\stackrel{\stackrel{\rm not}{t-{\rm like}}}{\sigma^2}} {\pmpi_{\sigma^2}\circ
R_{\sigma^2}(\,^{\pm}\!\Omega)}\right )}\prod_{\stackrel{\stackrel{\rm
 not}{t-{\rm like}}}{\sigma^2}}{\rm d}^3
\pmbpi_{\sigma^2}\prod_{\sigma^3}{{\cal D}\,^{\pm}\!\Omega_{\sigma^3}}.
\end{eqnarray}

\noindent The group property SO(4) = SU(2) $\!\times\!$ SU(2) is used by decomposing
generator $w$ of the connection $\Omega$ = $\exp w$ as well as any antisymmetric
matrix into self- and antiselfdual parts,
\begin{equation}                                                                    
w = \,^+\!w + \,^-\!w, ~~~ {1\over 2}\epsilon^{ab}{}_{cd}\,^\pm\!w^{cd} = \pm
\,^\pm\!w^{ab}.
\end{equation}

\noindent Correspondingly, $\Omega$ as well as $R$ are decomposed multiplicatively,
\begin{equation}                                                                    
\label{Omega-dual}%
\Omega = \,^+\!\Omega \,^-\!\Omega, ~~~ \,^\pm\!\Omega = \exp \,^\pm\!w.
\end{equation}

\noindent The basis of (anti-)selfdual matrices $\pmSig^k_{ab}$ ($k$ = 1, 2, 3) is
introduced such that $i\pmSig^k_{ab}$ satisfy Pauli matrices algebra. Thereby
(anti-)selfdual parts of a tensor $v_{ab}$, the $\pmv_{ab}$, are parameterised by the
3-vector components,
\begin{equation}                                                                    
\pmv_{ab} = {1\over 2}\pmv_k\pmSig^k_{ab} = {1\over 2}\pmbv\pmbSig_{ab}
\end{equation}

\noindent (so that
\begin{equation}                                                                    
|v|^2 \stackrel{\rm def}{=} v\circ v = {1\over 2}|\pbv|^2 + {1\over 2}|\mbv|^2
\end{equation}

\noindent ) as well as generators of the connections and curvatures,
\begin{equation}                                                                    
\,^{\pm}\!\Omega = \exp (\pmw_k\pmSig^k), ~~~ \,^{\pm}\!R = \exp
(\,^{\pm}\!\phi_k\pmSig^k).
\end{equation}

\noindent The latter in the given notations correspond to the SU(2) rotations by the
angles $|\pmbw|$, $|\,^{\pm}\!\bphi|$. At the same time, these matrices can be
considered in the adjoint representation, that is, as those acting on the vectors
$\pmv_k$ as SO(3) rotations by twice as much angles, for example,
\begin{equation}                                                                    
\exp (\pmw_k\pmSig^k)\pmv_l\pmSig^l\exp (-\pmw_m\pmSig^m) = \left ({\cal
O}(2\pmbw)\pmbv\right )_k\pmSig^k
\end{equation}

\noindent is rotation by the angle $2|\pmbw|$ around the axis $\pmbw$.

In addition to the measure (\ref{d-mu+-}), consider also the versions
\begin{eqnarray}                                                                   
\label{d-mu+-SU(2)}%
{\rm d} \mu^{\rm SU(2)}_{\rm area} & = & {\rm d} \,^+\!\mu^{\rm SU(2)}_{\rm area}{\rm
d} \,^-\!\mu^{\rm SU(2)}_{\rm area}, \nonumber\\ {\rm d} \,^{\pm}\!\mu^{\rm
SU(2)}_{\rm area}(-i\{\pi\},\{\Omega\}) & = & \exp{\left (-\! \sum_{\stackrel{t-{\rm
like}}{\sigma^2}}{|\pmbtau_{\sigma^2}|\arcsin{\pmtau _{\sigma^2}\circ
R_{\sigma^2}(\,^{\pm}\!\Omega)\over |\pmbtau_{\sigma^2}|}}\right )}\nonumber\\
 & & \hspace{-40mm} \exp{\left (i
\!\sum_{\stackrel{\stackrel{\rm not}{t-{\rm like}}}{\sigma^2}}
{|\pmbpi_{\sigma^2}|\arcsin{\pmpi_{\sigma^2}\circ R_{\sigma^2}(\,^{\pm}\!\Omega)\over
|\pmbpi_{\sigma^2}|}}\right )}\prod_{\stackrel{\stackrel{\rm
 not}{t-{\rm like}}}{\sigma^2}}{\rm d}^3
\pmbpi_{\sigma^2}\prod_{\sigma^3}{{\cal D}\,^{\pm}\!\Omega_{\sigma^3}}
\end{eqnarray}

\noindent or
\begin{eqnarray}                                                                   
\label{d-mu+-SO(3)}%
{\rm d} \mu^{\rm SO(3)}_{\rm area} & = & {\rm d} \,^+\!\mu^{\rm SO(3)}_{\rm area}{\rm
d} \,^-\!\mu^{\rm SO(3)}_{\rm area}, \nonumber\\ {\rm d} \,^{\pm}\!\mu^{\rm
SO(3)}_{\rm area}(-i\{\pi\},\{\Omega\}) & = & \exp{\left (-{1\over 2}\!
\sum_{\stackrel{t-{\rm like}}{\sigma^2}}{|\pmbtau_{\sigma^2}|\arcsin{\pmbtau
_{\sigma^2} * R_{\sigma^2}(\,^{\pm}\!\Omega)\over |\pmbtau_{\sigma^2}|}}\right
)}\nonumber\\
 & & \hspace{-40mm} \exp{\left ({i\over 2}
\!\sum_{\stackrel{\stackrel{\rm not}{t-{\rm like}}}{\sigma^2}}
{|\pmbpi_{\sigma^2}|\arcsin{\pmbpi_{\sigma^2} * R_{\sigma^2}(\,^{\pm}\!\Omega)\over
|\pmbpi_{\sigma^2}|}}\right )}\prod_{\stackrel{\stackrel{\rm
 not}{t-{\rm like}}}{\sigma^2}}{\rm d}^3
\pmbpi_{\sigma^2}\prod_{\sigma^3}{{\cal D}\,^{\pm}\!\Omega_{\sigma^3}},
\end{eqnarray}

\noindent where $\,^{\pm}\!\Omega$, $R(\,^{\pm}\!\Omega)$ are SU(2) or SO(3)
rotations, respectively. The arguments in favour of the measures (\ref{d-mu+-SU(2)}),
(\ref{d-mu+-SO(3)}) are given in our earlier work \cite{Kha3}: these measures still
can be defined to satisfy the canonical quantisation principle now in addition to the
correspondence principle (proportionality to $e^{-S}$ due to validity of the selfdual
representation of Regge action \cite{Kha5}). This is due to a possibility to give
sense to the 3-dimensional integral
\begin{equation}                                                                   
\label{g-represent-delta}%
\int{e^{\textstyle ilg(R*\bl/l)}{\d^3\bl\over (2\pi)^3}}, ~~~ g(x) = -g(-x), ~~~
g^{\prime}(0) = 1,
\end{equation}

\noindent so that it does not depend on the details of behaviour of the function
$g(x)$ analytical in the neighbourhood of zero. Namely, let us pass to spherical
coordinates so that $\d^3\bl$ = $l^2\d l$ $\!\d o_{\sbl}$, divide integral over the
angles $\d o_{\sbl}$ into those ones over the upper and over the lower hemispheres and
represent the latter as the integral over the upper hemisphere and over negative $l$,
\begin{eqnarray}                                                                   
\label{int-g-to-delta}%
\lefteqn{\int{e^{\t ilg(r\cos\theta) - \ve l^2}l^2\d l\sin\theta\d\theta\d\varphi} =
<\cos\theta = z>} \nonumber\\ & = & 2\pi\int\limits^1_0\d z\int\limits^{\infty}_0e^{\t
ilg(rz) - \ve l^2}l^2\d l + 2\pi\int\limits^0_{-1}\d z\int\limits^{\infty}_0e^{\t
ilg(rz) - \ve l^2}l^2\d l \nonumber\\ & = & 2\pi\int\limits^1_0\d
z\int\limits^{+\infty}_{-\infty}e^{\t ilg(rz) - \ve l^2}l^2\d l =
-(2\pi)^2\int\limits^1_0\delta^{\prime\prime}_{\ve}\left (g(rz)\right )\d z
\nonumber\\ & = & -{1\over
2}(2\pi)^2\int\limits^1_{-1}\delta^{\prime\prime}_{\ve}\left (g(rz)\right )\d z ~~~
\stackrel{\ve\to 0}{\longrightarrow} ~~~ -{1\over
2}(2\pi)^2\int\limits^1_{-1}\delta^{\prime\prime}\left (g(rz)\right )\d z,
\end{eqnarray}

\noindent where $r$ = $|\br|$, $r^a$ $\sdef$ $\epsilon^a{}_{bc}R^{bc}/2$. Here
\begin{equation}                                                                   
\delta\left (g(rz)\right ) = {\delta (rz)\over g^{\prime}(rz)} = \delta (rz),
\end{equation}

\noindent and a parameter $\ve$ $>$ 0 is introduced for temporary regularization of
the $\delta$-function to carefully take into account the integration limit $z$ = 0
(the edge of the hemispheres) of the intermediate integrations which belongs to the
support of the $\delta$-function. Thus, for certain way of calculation the
conditionally convergent integral (\ref{g-represent-delta}) does not depend on the
details of behaviour of the function $g(x)$ with the specified properties and
therefore is equal to such the integral also at $g(x)$ = $x$,
\begin{equation}                                                                   
\label{represent-delta}%
\int{e^{\textstyle i\bl*R}{\d^3\bl\over (2\pi)^3}} = \delta^3\left ({R-\R\over
2}\right ),
\end{equation}

\noindent that is, it presents an alternative representation of the $\delta$-function.
This representation allows to raise the constraints from $\delta$-functions to
exponent in the Feynman path integral constructed in the canonical quantisation
formalism and to include them into the Lagrangian with the help of the Lagrange
multipliers over which integrations in the measure arise. Since a part of constraints
are just the antisymmetrical parts of curvature matrices, this can be used at $g(x)$ =
$\arcsin x$ in order to modify the construction of \cite{Kha1,Kha2} and to get Regge
action in the exponential. To get the 3-dimensional integrals of the type of
(\ref{g-represent-delta}) in four dimensions it is necessary to use (anti)selfdual
representations (as effectively 3-dimensional ones).

When calculating with the help of the measure obtained we should use the same
transition, but in backward direction, from (\ref{g-represent-delta}) to
(\ref{represent-delta}). Exactly in the same way the analogous expressions with the
product of the edge or area vectors,
\begin{equation}                                                                   
\label{v-g-deltaSU(2)}%
\int e^{\t i|\pmbv|g(\,^{\pm}\!R\circ\pmv/|\pmbv|)}
\pmv^{k_1}\pmv^{k_2}\ldots\pmv^{k_n} {\d^3\pmbv\over (2\pi)^3}, ~~~ g(x) = -g(-x), ~~~
g^{\prime}(0) = 1,
\end{equation}

\noindent or
\begin{equation}                                                                   
\label{v-g-deltaSO(3)}%
\int e^{\t i|\pmbv|g(\,^{\pm}\!R * \pmbv/|\pmbv|)/2}
\pmv^{k_1}\pmv^{k_2}\ldots\pmv^{k_n} {\d^3\pmbv\over (2\pi)^3}, ~~~ g(x) = -g(-x), ~~~
g^{\prime}(0) = 1,
\end{equation}

\noindent do not depend on the details of behaviour of an analytic function $g(x)$ and
reduce to the derivatives of the $\delta$-functions $\delta^3(\,^+\!R$ $-$
$\!^+\!\bar{R})$, $\delta^3(\,^-\!R$ $-$ $\!^-\!\bar{R})$.

In particular, the integrals in (\ref{d-mu+-SU(2)}), (\ref{d-mu+-SO(3)}) in the
degenerate case $\tau$ = 0 can be defined in such the way as if we would replace
$\arcsin x$ $\to$ $x$ in all the terms there. The difference of the measure
(\ref{d-mu+-SU(2)}) from the analogous one without 'arcsin', (\ref{d-mu+-}), should
display explicitly in calculations at $\tau$ $\neq$ 0, which are just made for the
simplest configurations of the area tensor Regge calculus below.

After these preliminaries let us proceed with calculation.

For definiteness, consider Regge lattice composed of the two identical, up to
reflection, 4-simplices with the vertices 0, 1, 2, 3, 4. We can imagine this complex
if consider the two 4-simplices (01234) and $(0^{\prime}1234)$ with common 3-face
(1234) and identify the vertices 0 and $0^{\prime}$. Let all the connections $\Omega$
on the 3-faces act from (01234) to $(0^{\prime}1234)$, that is, if a 2-face tensor
$\pi$ is defined in (01234), then $\Omega\pi\bar{\Omega}$ is defined in
$(0^{\prime}1234)$. Denote a 3-face in the same way as the opposite vertex, $\Omega_i$
$\equiv$ $\Omega_{\sigma^3}$ where $\sigma^3$ = $(\{01234\}\setminus \{i\})$. Here
\{\dots\} denote (sub)set, here of the vertices 0, 1, 2, 3, 4. Denote a 2-face in the
same way as the opposite edge, $v_{(ik)}$ $\equiv$ $v_{\sigma^2}$, $R_{(ik)}$ $\equiv$
$R_{\sigma^2}$ where $\sigma^2$ = $(\{01234\}\setminus \{ik\})$. It is convenient to
define the variables $v$, $R$ on the ordered pairs of vertices $ik$, $v_{ik}$ =
$-v_{ki}$, $R_{ik}$ = $\bar{R}_{ki}$. (Then $v_{(ik)}$ is a one of the two values,
$v_{ik}$ or $v_{ki}$, $R_{(ik)}$ is $R_{ik}$ or $R_{ki}$). The action takes the form
\begin{equation}                                                                   
S(v,\Omega) = \sum^4_{i<k}{|v_{ik}|\arcsin{v_{ik}\circ R_{ik}\over |v_{ik}|}},
~~~R_{ik} = \bar{\Omega}_i\Omega_k.
\end{equation}

The curvature matrices are related by the Bianchi identities $R_{ik}R_{kl}$ = $R_{il}$
(on the triangles with common edge $(\{01234\}\setminus \{ikl\})$). As independent
curvature matrices we can choose $R_{\alpha}$ $\equiv$ $R_{0\alpha}$ ($\alpha$,
$\beta$, $\gamma$, \dots = 1, 2, 3, 4), that is, the curvature on the 2-faces of the
tetrahedron (1234). Thereby, in accordance with our approach \cite{Kha1}, the
tetrahedron (1234) can be attributed to the leaf while the other triangles
$(0\alpha\beta)$ can be treated as $t$-like ones. This corresponds to the naive
viewpoint on this Regge lattice as that one corresponding to the evolution in time of
the closed manifold consisting of the two copies of the terahedron (1234) to the point
0. Upon dividing the triangles into the $t$-like ones with the tensors
$\tau_{\alpha\beta}$ $\equiv$ $v_{\alpha\beta}$ entering as parameters and the leaf
ones with the tensors $\pi_{\alpha}$ $\equiv$ $v_{0\alpha}$ the action reads
\begin{equation}                                                                   
S(v,\Omega) = \sum^4_{\alpha = 1}{|\pi_{\alpha}|\arcsin{\pi_{\alpha}\circ
R_{\alpha}\over |\pi_{\alpha}|}} + \sum^4_{\alpha < \beta
}{|\tau_{\alpha\beta}|\arcsin{\tau_{\alpha\beta}\circ (\bar{R}_{\alpha}R_{\beta})\over
|\tau_{\alpha\beta}|}}.
\end{equation}

Consider averaging a function of a single area tensor, for example, of $\pi_1$ with
the help of the measure (\ref{VEV2}) (or (\ref{d-mu+-})). Integration over $\d^6\pi_2$
$\!\d^6\pi_3$ $\!\d^6\pi_4$ yields $\delta$-functions of (the antisymmetric parts of)
the curvatures which being integrated over ${\cal D}R_2$ $\!{\cal D}R_3$ $\!{\cal
D}R_4$ are reduced to unity. As a result, the measure on the functions of $\pi_1$
follows which differs from the case $\tau$ = 0 \cite{Kha1} by shifting by nonzero
average $<\pi_1>$,
\begin{eqnarray}                                                                   
\label{<f(pi1)>}%
<f(\pi_1)> & = & \int f(-i\pi_1)\d^6\pi_1 \int e^{\t i(\pi_1 - <\pi_1>)\circ R}{\cal
D}R\nonumber\\ & = & \int f(\pi_1)\Theta (\pi_1 - <\pi_1>) {\d^3\pbpi_1\over 4\pi}
{\d^3\mbpi_1\over 4\pi},\nonumber\\ \Theta (\pi_1) & = & {\nu_2(|\pbpi_1|)\over
|\pbpi_1|^2}{\nu_2(|\mbpi_1|)\over |\mbpi_1|^2}, ~~~ <\pi_1> = \sum^4_{\alpha =
2}\tau_{1\alpha}.
\end{eqnarray}

\noindent It is seen that $<\pi_1>$ is the value of the tensor of the triangle (234)
expected from the closure of the surface of the tetrahedron (0234) with known values
of the tensors $\tau$ of the other 2-faces (023), (034), (042).

The result (\ref{<f(pi1)>}) holds also for an arbitrary structure of the Regge lattice
if integration over the rest of tensors $\pi$ is made. The area tensor $\pi_1$
fluctuate around the expectation $<\pi_1>$ with Plank scale of the fluctuations. On
the Regge lattice of a regular periodic structure, the $<\pi>$ is defined from the
closure condition of some 3-dimensional prism whose lateral surface is formed by the
$t$-like triangles. The $\pi_1$ is the tensor of one of the bases of this prism.
Another base is a leaf or diagonal triangle as well, the tensor of which
$\pi^{\prime}_1$ certainly enters the closure condition. Therefore $\pi^{\prime}_1$
should be given as initial condition. On the whole, the area tensors $\pi$ are to be
fixed on some 3-dimensional section consisting of the leaf or diagonal triangles.

Now let us use the measure ${\rm d} \mu^{\rm SU(2)}_{\rm area}$ (\ref{d-mu+-SU(2)}) to
average $f(\pi_1)$. According to the above method of giving sense to the integral of
exponent of $i|\pmbpi|\arcsin(\dots)$, the integrations over $\d^6\pi_2$ $\!\d^6\pi_3$
$\!\d^6\pi_4$ can be performed after the replacement $\arcsin x$ $\to$ $x$ and yield
$\delta$-functions of (the antisymmetric parts of) the curvatures which being
integrated over ${\cal D}R_2$ $\!{\cal D}R_3$ $\!{\cal D}R_4$ are reduced to unity. To
be able to get an exact result let us confine ourselves by the case when only one of
the tensors $\tau_{1\alpha}$ differs from zero, for example, $\tau_{14}$ $\equiv$
$\tau$,
\begin{eqnarray}                                                                   
\lefteqn{<f(\pi_1)>_{\rm SU(2)} =} \nonumber\\ & & \phantom{(\pi_1)>}\int \exp \left
(- |\pbtau|\arcsin{\,^+\!\tau\circ \,^+\!R\over |\pbtau|} -
|\mbtau|\arcsin{\,^-\!\tau\circ \,^-\!R\over |\mbtau|} \right ){\cal D}\,^+\!R{\cal
D}\,^-\!R\nonumber\\ \lefteqn{\int \exp \left
(i|\!\pbpi_1|\arcsin{\!\,^+\!\pi_1\!\circ \!\,^+\!R\over |\pbpi_1|} +
i|\!\mbpi_1|\arcsin{\!\,^-\!\pi_1\!\circ \!\,^-\!R\over |\mbpi_1|}\right
)\!f(-i\pi_1)\, \d^3\!\pbpi_1 \d^3\!\mbpi_1,}
\end{eqnarray}

\noindent and let the function $f(\pi_1)$ depends on only the scalars $|\pmbpi_1|$. In
the case of probe functions being the product of powers $[(\pmbpi_1)^2]^k$ we have
integrals
\begin{eqnarray}                                                                   
\label{<l>SU(2)}%
\lefteqn{\int{\sin^2{\!\phi }\over\phi^2}{\rm d}^3\bphi \,e^{\t -\at\arcsin
(\bn_{\sbtau}\bn_{\sbphi}\sin\phi)} \int e^{\t il\arcsin
(\bn_{\sbl}\bn_{\sbphi}\sin\phi)}(-l^2)^k\d^3\bl} \nonumber\\ & & = \int{\sin^2{\!\phi
}\over\phi^2}{\rm d}^3\bphi \left (\int\limits^1_{-1}{\d z\over 2}e^{\t -\at \arcsin
(z\sin\phi) }\right )\int e^{\t i\bl\bn_{\sbphi}\sin\phi} (-l^2)^k\d^3\bl.
\end{eqnarray}

\noindent Here we denote $\bl$ = $\pmbpi$, $\btau$ = $\pmbtau$, $l$ = $|\bl|$, and
$\bn_{\sbl}$, $\bn_{\ldots}$ are unit vectors in the directions $\bl$, \ldots; the
vectors $\bphi$ = $\pmbphi$ parameterise SU(2) rotations $\,^{\pm}\!R$. In accordance
with the above method of giving sense to the integral of a power of $\bl$ times
exponent of $i|\pmbpi|\arcsin(\dots)$, the substitution $\arcsin x$ $\to$ $x$ in the
integral over $\d^3\bl$ is made. Round brackets contain an integral over $\d z$ of the
expression with yet another $\arcsin$. It is some function of $\sin\phi$. The whole
integral for the latter being probe power function reads
\begin{equation}                                                                   
\int {\sin^2\!\phi \over \phi^2}\d^3 \!\bphi \,\sin^{2m}\!\phi \!\int e^{\t
i\bl\bn_{\sbphi} \sin \phi} (-\bl^2)^k {\d^3\bl \over (2\pi)^3} = (2k + 1)!{(2k - 2m -
1)!!\over 2^{k - m}(k - m)!}
\end{equation}

\noindent which can be readily extended to arbitrary function of $l^2$,
$\sin^2\!\phi$,
\begin{eqnarray}                                                                   
\lefteqn{\int {\sin^2\!\phi \over \phi^2}\d^3 \!\bphi \int e^{\t i\bl\bn_{\sbphi} \sin
\phi} F(-l^2, \sin^2\!\phi) \d^3\bl =} \nonumber\\ & & \phantom{\int {\sin^2\!\phi
\over \phi^2}\d^3 \!\bphi \int e^{\t i\bl\bn_{\sbphi} \sin \phi} F(-l^2}\int \d^3\bl
{2\pi \over l} \int\limits^{+\infty}_{-\infty} e^{\t -l\ch\alpha} F(l^2, \ch^2\alpha)
\ch\!\alpha\, \d\alpha.
\end{eqnarray}

\noindent As a result, the absolutely convergent integrals follow,
\begin{eqnarray}                                                                   
\label{<f(pi)>SU(2)}%
\lefteqn{<f(|\pbpi_1|, |\mbpi_1|)>_{\rm SU(2)} =} \nonumber\\ & & \phantom{\nu_2(\at,
l)}\int f(|\pbpi_1|, |\mbpi_1|) \nu_2(|\pbtau|, |\pbpi_1|) \nu_2(|\mbtau|, |\mbpi_1|)
\d |\pbpi_1| \d |\mbpi_1|, \nonumber\\ \lefteqn{\nu_2(\at, l) = {\ch {\t \pi\at\over
{\t 2}}\over 1 + \at^2} {l \over \pi} \int\limits^{+\infty}_{-\infty} e^{\t
-l\ch\alpha} \left [\cos (\alpha\at)\ch\alpha + \at\sin (\alpha\at)\sh\alpha \right
]\d\alpha.}
\end{eqnarray}

Averaging with the help of the analogously constructed measure, but now on the basis
of SO(3) (anti)selfdual rotations, ${\rm d} \mu^{\rm SO(3)}_{\rm area}$
(\ref{d-mu+-SO(3)}) on the same functions leads to the result differing from
(\ref{<f(pi)>SU(2)}) apart from the overall normalisation by only the factor $\ch^{-2}
\alpha$ under the integral sign in the formula for $\nu$,
\begin{eqnarray}                                                                   
\label{<f(pi)>SO(3)}%
\lefteqn{<f(|\pbpi_1|, |\mbpi_1|)>_{\rm SO(3)} =} \nonumber\\ & & \phantom{\nu_3}\int
f(|\pbpi_1|, |\mbpi_1|) \nu_3\left (|\pbtau|, {|\pbpi_1|\over 2}\right ) \nu_3\left
(|\mbtau|, {|\mbpi_1|\over 2}\right ) {\d |\pbpi_1|\over 2} {\d |\mbpi_1|\over 2},
\nonumber\\ \lefteqn{\nu_3(\at, l) =} \nonumber\\ & & {\ch {\t \pi\at\over {\t
2}}\over 1 + \at^2} {1\over 1 + {\t \at^2\over {\t 3}}} {2l \over \pi}
\int\limits^{+\infty}_{-\infty} e^{\t -l\ch\alpha} \,{\cos (\alpha\at)\ch\alpha +
\at\sin (\alpha\at)\sh\alpha \over \ch^2\!\alpha }\d\alpha.
\end{eqnarray}

The results (\ref{<f(pi)>SU(2)}) and (\ref{<f(pi)>SO(3)}) for expectation of a
function of $\pi_1$ hold also for an arbitrary structure of the Regge lattice when
only one of the tensors $\tau$ is nonzero and integration over the tensors $\pi$ other
than $\pi_1$ is performed.

The trigonometrical function $\cos (\alpha\at)$ or $\sin (\alpha\at)$ in the integrals
for $\nu$ is modulated by a positive function decreasing at infinity. If the latter is
not monotonic, it may be the case at small $l$ that contribution from the negative
half-period of the trigonometric function can dominate contribution from the positive
half-period. The $\nu_2 (\at, l)$ is reduced to a combination of the modified Bessel
functions of complex orders with corresponding asymptotic at $l$ $\to$ 0 and fixed
$\at$,
\begin{eqnarray}                                                                   
\nu_2(\at, l) & = & {\ch {\t \pi\at\over {\t 2}}\over 1 + \at^2} {2l \over \pi} {\rm
Re} [(1 - i\at)K_{1 + i\sat}(l)] \nonumber\\ & \sim & {2\over \pi} {\ch {\t
\pi\at\over {\t 2}}\over \sqrt{1 + \at^2}} \sqrt{\pi\at \over \sh (\pi\at)} \cos\left
[\at\ln {2e^{1+\gamma}\over l} + O(\at^3)\right ],
\end{eqnarray}

\noindent where $\gamma$ = 0.5772\dots (the Euler constant). At $$l < 2\exp \left
[-{\pi\over 2\at} + 1 + \gamma + O(\at^2)\right ]$$ the $\nu_2$ becomes variable in
sign.

Negativity of $\nu_2 (\at, l)$ in some intervals of the values of $l$ does not prevent
from probabilistic interpretation at such $l$ of the full measure on the physical
surface because the full measure contains the product of the two $\nu$'s, on selfdual
and antiselfdual sectors. Therefore the measure is positive as the square of $\nu$ if
being reduced to the physical surface where $|\pbv|$ = $|\mbv|$ what has been done in
our work \cite{Kha4}. Negativity of $\nu_2$ is essential when trying to give
probabilistic interpretation of the measure also outside the physical surface. This is
somewhat philosophical problem.

Contrary to $\nu_2(\at, l)$, the $\nu_3(\at, l)$ is positive for all $l$ if $\at$ is
bounded from above by the value of the order of Plank scale. To show this, use the
following fact. The integral
\begin{equation}                                                                   
\label{f-cos-x}%
\int\limits^{\infty}_0 f(x) \cos x\,\d x, ~~~ f^{\prime\prime}(x) > 0 ~~ (0 < x <
\infty)
\end{equation}

\noindent is positive ($f^{\prime\prime}(x)$ $>$ 0 and convergence of the integral
mean that $f(x)$ $>$ 0, $f^{\prime}(x)$ $<$ 0 on the whole interval 0 $<$ $x$ $<$
$\infty$). This is easily seen upon dividing the integration region into the periods
$2\pi$ and equivalently rewriting integral in each period as follows,
\begin{eqnarray}                                                                   
\lefteqn{\int\limits^{\infty}_0 f(x) \cos x\,\d x = \sum^{\infty}_{n=0}
\int\limits^{\pi}_0 \left\{ \left [f_n(x) - f_n\left ({\pi\over 2}\right )\right ]
\right.} \nonumber\\ & & \phantom{\int\limits^{\infty}_0 f(x)}\left. - \left [f_n(x +
\pi) - f_n\left ({\pi\over 2} + \pi\right ) \right ] \right\} \cos x\,\d x, ~~~ f_n
(x) \stackrel{\rm def}{=} f(x + 2\pi n).
\end{eqnarray}

\noindent Since $f^{\prime\prime}(x)$ $>$ 0 (the function is concave), the expression
in braces at 0 $<$ $x$ $<$ ${{\t \pi}\over {\t 2}}$ is positive and at ${{\t \pi}\over
{\t 2}}$ $<$ $x$ $<$ $\pi$ is negative, just as $\cos x$. This proves positivity of
(\ref{f-cos-x}). As for the expression for $\nu_3(\at, l)$,
\begin{equation}                                                                   
\int\limits^{\infty}_0 [f_1(\alpha )\cos (\alpha\at ) + f_2(\alpha )\sin (\alpha\at
)]\,\d\alpha,
\end{equation}

\noindent we have, first, $f^{\prime}_1(0)$ = 0 for even function so that
$f^{\prime\prime}_1(\alpha)$ $<$ 0 in some neighbourhood of $\alpha$ = 0. Second,
there is also the term with $\sin (\alpha\at)$. Therefore integrating by parts reduce
the triginometric functions to $\sin (\alpha\at)$ and divide integral into those ones
over the regions (0, $\pi / (2\at))$ and $(\pi / (2\at)$, $\infty )$. Redenoting
$\alpha\at$ $\to$ $\alpha\at$ + $\pi / 2$ in the latter region we get an integral over
(0, $\infty )$ with $\cos (\alpha\at)$,
\begin{equation}                                                                   
\int\limits^{\infty}_0 h(\alpha )\sin (\alpha\at )\,\d\alpha =
\int\limits^{\pi/(2\sat)}_0 h(\alpha )\sin (\alpha\at )\,\d\alpha +
\int\limits^{\infty}_0 h\left (\alpha + {\pi\over 2\at}\right )\cos (\alpha\at
)\,\d\alpha
\end{equation}

\noindent where $h(\alpha)$ = $-f^{\prime}_1(\alpha)/\at$ or $f_2(\alpha)$. The first
integral here is positive, the second one is of the type of (\ref{f-cos-x}) if
$h^{\prime\prime}(\alpha)$ $>$ 0 at $\alpha$ $>$ $\pi/(2\at)$. That is, if
$h^{\prime\prime}(\alpha)$ $>$ 0 $\forall$ $\alpha$ $>$ $\alpha_0$ then at $\at$ $<$
$\pi/(2\alpha_0)$ positivity is guaranteed. The additional factor $\ch^{-2}\alpha$ in
the integral for $\nu_3(\at, l)$ as compared to $\nu_2(\at, l)$ provides the
expression under the integral sign to decrease at $\alpha$ $\to$ $\infty$ also for $l$
= 0, therefore the estimate of $\alpha_0$ can be made independent of $l$ (in the case
of $\nu_2(\at, l)$ $\alpha_0$ $\to$ $\infty$ at $l$ $\to$ 0). Elementary analysis of
the graph of the function $h(\alpha)$ = $-f^{\prime}_1(\alpha)/\at$ yields
$h^{\prime\prime}(\alpha)$ $>$ 0 at $\ch^2\alpha$ $>$ $\ch^2\alpha_0$ = $y_0$ =
6.1108\dots where $y_0$ is the root of the cubic equation
\begin{equation}                                                                   
(y - 6)\left [y^2 + {4\over 27}(y - 6)(y - 1) - {4\over 3}\right ] = 4.
\end{equation}

\noindent Such $\alpha$ ensures also $f^{\prime\prime}_2(\alpha)$ $>$ 0 since it is
sufficient to have $\ch^2\alpha$ $>$ 6 to provide this inequality uniformly in $l$. As
a result, if
\begin{equation}                                                                   
\at \leq {\pi\over 2\alpha_0} = {\pi\over 2\ln (\sqrt{y_0} + \sqrt{y_0 - 1})} = 1.010
\dots,
\end{equation}

\noindent the positivity of $\nu_3(\at, l)$ is guaranteed. In the case of $\nu_2(\at,
l)$ we would have the estimate $\at$ $<$ $\pi/[2\ln (2e^{1 + \gamma}/l)]$ +
$O(\ln^{-4}(2e^{1 + \gamma}/l))$, that is, we need $\at$ $\to$ 0 to ensure positivity
of $\nu_2(\at, l)$ at $l$ $\to$ 0.

Thus, the measures (\ref{d-mu+-}), (\ref{d-mu+-SU(2)}) and (\ref{d-mu+-SO(3)}) lead to
reasonable consequences for vacuum expecta\-tions. All these have smooth limit
$|\pmbtau|$ $\to$ 0 in which (\ref{d-mu+-}) and (\ref{d-mu+-SU(2)}) coincide. The
measure (\ref{d-mu+-SU(2)}) based on SU(2) rotations has zones at small $|\pmbpi|$ (in
Plank scale) where it is negative outside the physical surface on which $|\pbv|$ =
$|\mbv|$ in the configuration space of arbitrary independent area tensors. The measure
(\ref{d-mu+-SO(3)}) based on SO(3) rotations is expected to be positive also at
arbitrary independent area variables $\pmbpi$, not only on the physical surface, if we
restrict the vectors $|\pmbtau|$ of the $t$-like triangles in absolute value by a
number of the order of Plank scale.

\bigskip

The present work was supported in part by the Russian Foundation for Basic Research
through Grant No. 03-02-17612.


\begin{thebibliography}{99}
\bibitem{Kha1}
 V.M. Khatsymovsky, Phys. Lett. {\bf 560B}, 245 (2003), gr-qc/0212110.
\bibitem{Kha2}
 V.M. Khatsymovsky, Class. Quantum Grav. {\bf 11}, 2443 (1994), gr-qc/9310040.
\bibitem{Wael}
 H. Waelbroeck, Class. Quantum Grav. {\bf 7}, 751 (1990).
\bibitem{MisThoWhe}
 C.W. Misner, K.S. Thorne, J.A. Wheeler, Gravitation. W.H. Freeman and Company, San
 Francisco, 1973.
\bibitem{Kha3}
 V.M. Khatsymovsky, Feynman path integral in area tensor Regge calculus and
 cor\-respondence principle, gr-qc/0406049.
\bibitem{Kha5}
 V.M. Khatsymovsky, Class. Quantum Grav. {\bf 6}, L249 (1989).
\bibitem{Kha4}
 V.M. Khatsymovsky, Phys. Lett. {\bf 586B}, 411 (2004), gr-qc/0401053.
\end{thebibliography}
\end{document}